\documentclass[11pt]{article}
\usepackage[margin=1.25in]{geometry}                
\usepackage{graphicx}
\usepackage{amssymb}
\usepackage{epstopdf}
\usepackage{setspace}
\usepackage{amsmath}
\usepackage{multirow}
\usepackage{rotating}
\usepackage{algorithm}
\usepackage{enumerate}
\usepackage{url}
\usepackage{endnotes}
\usepackage[usenames,dvipsnames]{color}

\usepackage[sort]{natbib}

\newcommand{\citepos}[1]{}

\renewcommand{\citepos}[1]{\citeauthor{#1}'s (\citeyear{#1})}

\title{Contending Parties: A Logistic Choice Analysis of Inter- and Intra-group Blog Citation Dynamics in the 2004 US Presidential Election\thanks{This work was supported in part by grants from US ONR \# N00014-08-1-1015, NSF \# OIA-102839, and NIH \# 1R01HD068395-01.}}
\author{Zack W. Almquist\footnote{Corresponding author: Department of Sociology; 3151 Social Science Plaza A; University of California, Irvine; Irvine, CA 92697-5100; {\tt almquist@uci.edu}} \and Carter T. Butts\footnote{Department of Sociology and Institute for Mathematical Behavioral Sciences; University of California, Irvine; {\tt buttsc@uci.edu}} }
\date{04/24/2011}

\begin{document}
\maketitle

\begin{abstract}

The 2004 US Presidential Election cycle marked the debut of Internet-based media such as blogs and social networking websites as institutionally recognized features of the American political landscape. Using a longitudinal sample of all DNC/RNC-designated blog-citation networks we are able to test the influence of various strategic, institutional, and balance-theoretic mechanisms and exogenous factors such as seasonality and political events on the propensity of blogs to cite one another over time. Capitalizing on the temporal resolution of our data, we utilize an autoregressive network regression framework to carry out inference for a logistic choice process. Using a combination of deviance-based model selection criteria and simulation-based model adequacy tests, we identify the combination of processes that best characterizes the choice behavior of the contending blogs.

\end{abstract}


\clearpage

\section{Introduction}

\doublespace

The 2004 US Presidential Election cycle marked the debut of Internet-based media such as blogs and social networking websites as institutionally recognized features of the American political landscape.  Particularly significant was the credentialing of selected blogs as officially designated media sources for purposes of covering the major political party conventions, an act which gave particular legitimacy to two contending groups of partisan blogs (one credentialed for the Republican National Convention (RNC) and the other for the Democratic National Convention (DNC)).  In the months that followed, these blogs served as significant foci for online journalistic, promotional, fund-raising, and organizing activities relating to the 2004 election.

In this study, we employ a dynamic logistic choice model to study the dynamics of interaction within and between these two groups of political blogs.  Using a longitudinal sample of all DNC and RNC-designated blog citation networks (sampled at six hour intervals for approximately four months) from \cite{butts09b} we are able to test for the influence of various strategic, institutional, and balance-theoretic mechanisms -- as well as exogenous factors such as seasonality and political events -- on the propensity of blogs to cite (i.e., hyperlink to) one another over time.  Capitalizing on the temporal resolution of our data, we utilize an autoregressive network regression framework to carry out inference for a logistic choice process closely related to the actor-oriented framework of \cite{snijders01}. 

This paper is structured as follows.  We begin by providing some general background from the relevant sociological and social network literatures, with a particular focus on the role of political blogs during the study period.  This is followed by a description of the study data, and an overview of our modeling approach. The latter includes both a discussion of the general assumptions behind the modeling of blog evolution as a dynamic decision process, and a treatment of the factors potentially shaping actors' payoffs.  We follow this with a discussion of our implementation and inferential framework, data analysis, and findings.  Finally, we conclude with a discussion of the implications of our results for our understanding of the social mechanisms shaping contentious groups in the online environment.

\section{Background}  

In recent years, the online world has generated a diverse array of new media for social interaction \citep{wellman01}, one of the most successful of which is the weblog (or ``blog'').  While a relatively obscure medium for many years, the growing popularity of blogs as a means for information dissemination, coordination, and political organization through the early to mid 2000s eventually led to their recognition of and adoption by established institutions.  A key landmark in this process was the 2004 US Presidential election cycle, in which the DNC and RNC first granted press credentials to selected bloggers for coverage of their national political conventions \citep{butts09b,adamic05,howard05,rainie05}.  This institutionalized legitimation by the major US political parties constituted a de facto recognition of the role of blogs (and the online community more broadly) as a durable element of the political landscape, and arguably marked the debut of the ``new media'' as a force in electoral politics.

The impact of blogs first gained institutional attention in the US political sphere in the early phases of the 2004 US electoral cycle, when Democratic presidential candidate and Vermont Governor Howard Dean rose to prominence partially as a result of his extensive use of online organizing to compensate for limited conventional resources in garnering media attention and raising funds \citep{ammori05,kerbel05}.  Dean's success in utilizing online interaction to mobilize a widely dispersed base of supporters was quickly noted by political observers, and (despite his loss of the Democratic nomination to Senator John Kerry) paved the way for other politicians to incorporate online media into their political campaigns \citep{cone03}.  Indeed, by the the end of the 2004 electoral cycle blogs and other online resources had been adopted by a number of Presidential contenders, and (via actions such as the above-mentioned credentialing of bloggers as members of the press) by the major US political parties themselves.  These and and further developments in the historical evolution of the online environment over the past decade have set the stage for academic, governmental, non-profit, and for-profit interest in blogs and other new media, particularly in political contexts \citep{drezner08}.

With respect to the role played by blogs per se, \cite{woodly08} demonstrates that blogs are actively used in mobilizing opinions, setting agendas, and generally influencing the elite members of the political parties.  His work demonstrates that the interactions between political blogs are a particularly important dimension of this phenomenon.  Because a distinctive feature of blogs is their combination of commentary on current events with hypertext references to primary or secondary information sources, the constantly evolving network of citations between blogs is at least as significant (e.g., from an information search standpoint) as the content of the individual blogs themselves.  Within this network of references, blog authors (or ``bloggers'') have become a new form of journalist, in some cases with similar information access and responsibilities to practitioners within traditional media outlets \citep{wall05}.  As the importance of this medium has continued to increase in recent years, its growth in size and elaboration has made its study both relevant and difficult.  We thus focus our attention on the initial ``watershed'' period of the 2004 US Presidential election, when the role of blogs as legitimated media entities was just beginning to crystalize.  In particular, our attention centers on the interactions among the relatively small number of blogs credentialed for the major party political conventions, as they jockeyed to promote their issues, candidates, and arguably themselves in the midst of a rapidly changing political and technological landscape.  As players with some institutional recognition but little control from established political actors, these blogs provide an early example of a phenomenon that has become increasingly common throughout the developed world.

\section{Data}

The data used in this paper is a dynamic inter- and intra-group blog citation network collected by \citet{butts09b}, consisting of interactions among all blogs credentialled by the DNC or RNC for their respective 2004 conventions.  Specifically, the set of actors (or vertex set) for this network consists of 34 DNC and 14 RNC credentialed blogs (with one blog credentialled by both groups) providing a combined network of 47 nodes observed over a 121 day period.  Network data was obtained by automatically querying the main page of each blog at six hour intervals starting at midnight, Pacific time.  The period of observation for this study begins on 7/22/04 (shortly before the DNC convention), and ends 11/19/04 (shortly after the Presidential election), leading to a total of 484 time points.  At each time point, the collected data consists of the network of URLs linking the main page of one blog to any page within another; i.e., there is an edge from blog $i$ to blog $j$ at time $t$ if a link to blog $j$ appears on the main page of $i$ at time $t$.  We may conceive of this data as an adjacency array, $A$, such that $A_{ij,t}=1$ if $i$ cites (i.e., links to) $j$ at time $t$, and 0 otherwise.  For purposes of this study, we ignore self-citations (i.e., internal links from a blog to itself).  

\begin{figure}[htbp]
   \centering
   \includegraphics[width=1\linewidth]{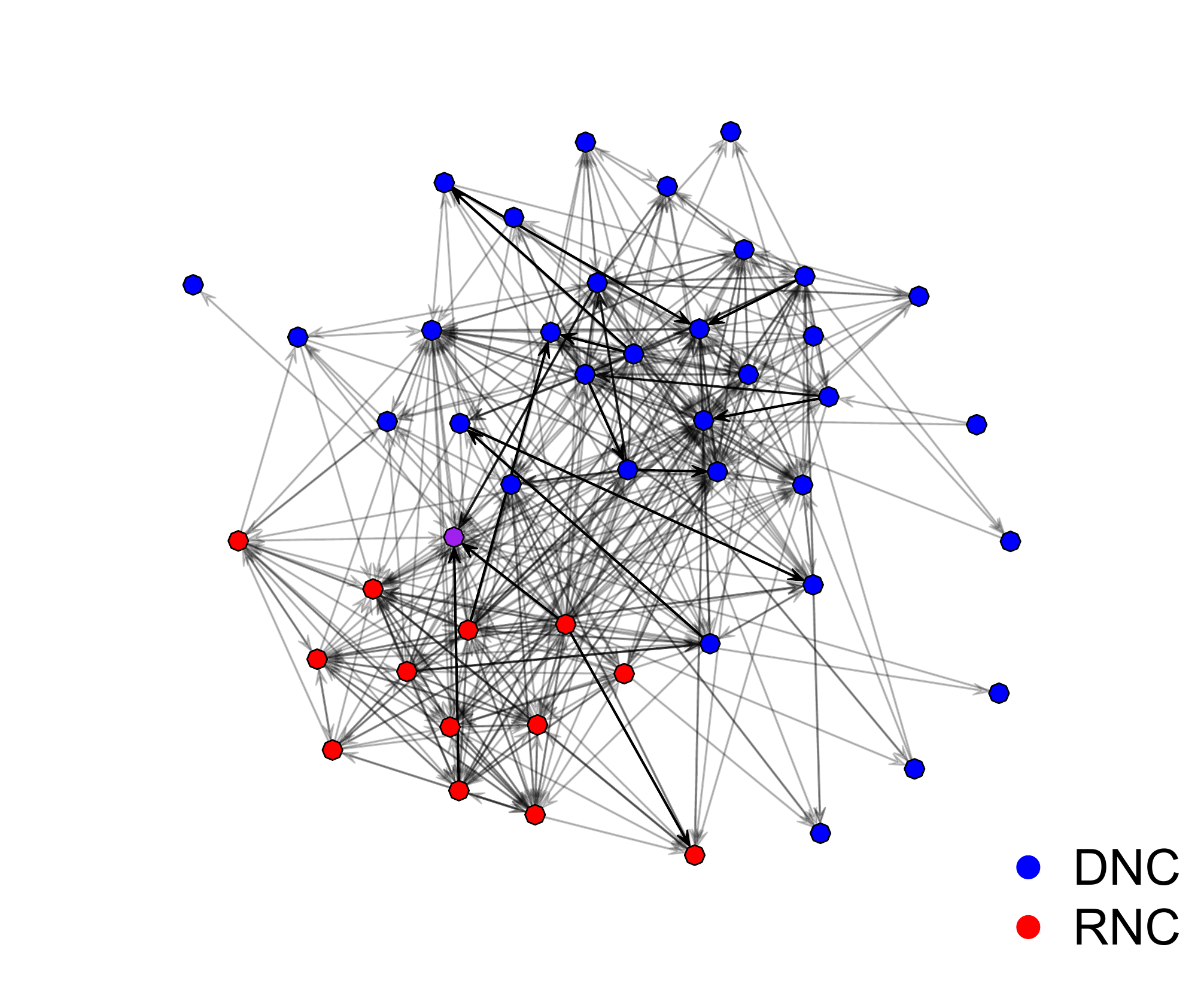}
   \caption{Combined RNC/DNC credentialled blog citation network aggregated over all time points, edges weighted by time-frequency. Purple node is credentialled by both the RNC and DNC.}
   \label{fig:comgraph}
\end{figure}

\begin{center}
[ Figure~\ref{fig:comgraph} About Here ]
\end{center}

In addition to the evolving blog network, \citet{butts09b} provide a timeline of major events during the campaign cycle, dividing the 121 day period into a series of ``epochs'' based on salient activities such as the RNC and DNC conventions, the televised Presidential debates, and the election itself (Table \ref{epochs}).  In an analysis of volatility within the RNC and DNC networks (taken separately), \citet{butts09b} find that these campaign events are related to the pace of change within the network (along with daily and weekly seasonal effects).  As such, we include these temporal effects as covariates in our analyses (as described below).

\begin{center}
[ Table \ref{epochs} About Here ]
\end{center}

\section{Network Evolution as a Decision Process} \label{sec:decision}
\label{sec:logitchoice}

Blogs of the type studied here are the deliberately constructed and maintained products of individuals, or small groups thereof.  Moreover, those blogs credentialed during the 2004 electoral cycle represented a small ``elite'' circle of especially active authors, whose blogs centered on coverage of politics and current events.  As such, it is reasonable to consider modeling the evolving blog network as arising from a dynamic decision process, in which blog authors select those to whom they link in response to context and past history.  This approach has been most fully developed by \citet{snijders96,snijders97} and \citet{snijders01}, who posit an ``actor-oriented'' model in which network members change their relationships via a latent continuous-time choice process.  We here employ a somewhat simpler version of this general scheme, which represents network evolution as a discrete time logistic choice process \citep{mcfadden76,mcfadden74}.  Although requiring somewhat stricter assumptions on decision simultaneity, this variant facilitates the accommodation of complex backward-looking behavior, and scales more easily to larger data sets.  

Although the inferential aspects of this framework will be described in Section~\ref{sec:method}, we begin here by presenting the model from a behavioral point of view.  First, we review the notion of edge updating as a logistic choice process \citep{snijders01}, with a specific emphasis on its interpretation in the present case.  As a revealed preference model, the logistic choice framework requires a parametric utility function; thus, we follow our initial discussion with a consideration of the payoff elements that may be expected to enter into blog authors' decision-making processes, as they decide to whom they will or will not link.  These payoff elements will form the core building blocks for our analysis of the evolving blog network.

\subsection{Edge Updating as Logistic Choice} \label{sec:logchoice}
\label{sec:edgeup}

At its crudest level, a blog is a web page with dynamically updated links to other online resources.  The core decision facing a blog author, then, is that of the other sites to which he or she should link, and (conversely) the links that can be removed (directly, or by allowing them to ``expire'' by no longer being shown on the blog's front page).  Such citations can be controlled on an individual basis, and are limited only by attentional and/or energetic costs: there is in principle no effective limit on the number of citations that can be maintained, and no barrier to adding or removing citations when desired.  At the same time, adding or removing links requires attention and effort on the part of the author, and is thus the result of deliberate action (as opposed, e.g., to the accidental, incidental, or automatic behaviors that are of considerable importance in face-to-face settings \citep{goffman59}).  Blog authors -- particularly active ones, such as those represented in this sample -- can and do spend considerable time monitoring their environment, and may thus be expected to be aware of and react to the actions of salient alters; moreover, recent citation history is relatively easily discovered in this environment, potentially facilitating the use of backward-looking strategies.  On the other hand, the complexity and dynamic nature of the online environment make prediction difficult, suggesting a very limited capacity for forward-looking behavior.

Taken together, the above considerations suggest the following propositions as a reasonable starting point for modeling the evolution of the blog network.  For simplicity of discussion, we will refer to the ``blog'' as the unit of decision making, and the links or citations from one blog to another as ``edges'' within the associated network.
\begin{enumerate}
\item The state of outgoing edges at each observation of the blog network is assumed to result from the choices of the sending blog;
\item Each blog in the network may send an edge to any number of other blogs in the network at any time;
\item The decision of a given blog regarding the state of a given edge is made myopically, and in isolation (i.e., the decision is considered on its own terms, without factoring in the effects of other decisions that might be made simultaneously);
\item The decision of a given blog regarding the state of a given edge may depend upon the past history of the blog network, or of the current external context (e.g., time of day, electoral cycle events)
\end{enumerate}
Subject to the above, we further presume that blog citation behavior follows a weakly consistent pattern of preferences, in the sense that there exists a \emph{utility function,} $u$, such that for the two alternative states $A_{ij,t}=0$ and $A_{ij,t}=1$, the odds that $i$ will choose $A_{ij,t}=1$ are strictly increasing in $u_i(A|A_{ij,t}=1)/u_i(A|A_{ij,t}=0)$.  Such a pattern of behavior is typically referred to as a \emph{stochastic choice} process, and can be viewed as a form of bounded rationality.  Although many stochastic choice models exist, we here use the common \emph{logistic choice} model.  In the present case, this amounts to the assumption that

\begin{equation}
\Pr(A_{ij,t}=1) = \frac{\exp\left[u_i\left(A |  A_{ij,t}=1\right)\right]}{\exp\left[u_i\left(A  |  A_{ij,t}=1\right)\right]+\exp\left[u_i\left(A  |  A_{ij,t}=0\right)\right]}, \label{eq:logit1}
\end{equation}
or, equivalently, that
\begin{equation}
\textrm{logit}\Pr(A_{ij,t=1}) = \ln\frac{\Pr(A_{ij,t}=1)}{\Pr(A_{ij,t}=0)} = u_i\left(A \: | \: A_{ij,t}=1\right)-u_i\left(A  | A_{ij,t}=0\right),\label{eq:logit2}
\end{equation}

i.e., the log-odds that $i$ will choose to cite $j$ at time $t$ is equal to the utility difference associated with sending (versus not sending) an edge.  Where the utility of one option is substantially greater than the other, then, actor behavior is nearly deterministic: the utility-increasing choice is selected with very high probability.  As the actor approaches indifference, however, choice behavior becomes increasingly random (an effect interpretable either as difficulty in determining the preferable option, or as reflecting the influence of various small, idiosyncratic payoffs).  When the actor is entirely indifferent between citing and not citing another, the choice becomes fully arbitrary (i.e., a coin flip).

To put this scheme into practice, we must make some further assumptions regarding the nature of the utility function.  From our list of propositions, we have assumed that decisions are made myopically, depending on the past (and on general context), but not on simultaneous or future decisions.  As such, we require that $u$ depend upon the network history, $A$, only through its prior states, and through the conjecturally perturbed state associated with a single decision (i.e., for the $A_{ij,t}$ decision, $u_i$ may depend upon $A_{\cdot\cdot, t-k}$ where $k>0$, and on $A_{\cdot\cdot, t}$ such that $A_{gh, t}=A_{gh, t-1}$ for all $g,h \neq i,j$).  $u$ may also depend upon $t$, and on exogenous covariates (denoted by $X$).  Finally, we will assume in general that $u$ can be written as a sum of linearly separable payoff elements, $s$, such that $u_i(A  |  A_{ij,t})=\theta^T s(A,A_{ij,t},i,j,t,X)$.  Intuitively, $s$ expresses the factors potentially driving $i$'s behavior, while the parameter vector $\theta$ expresses the direction and magnitude of the effect these factors have on the propensity to send or refrain from sending a tie.

As a model of boundedly rational dynamics, the logistic choice framework is quite general: a wide range of factors can potentially enter into the utility function, and the choice of possible candidates must be made based on substantive considerations.  With that in mind, we now turn to a consideration of the payoff elements that may plausibly drive behavior within the blog network.

\subsection{Potential Payoff Elements}
\label{hyp}

We apply three core hypotheses to the construction of the potential payoff elements that might influence this network. The first hypothesis is built around the long standing notion of preferential mixing \citep{mcpherson01} (e.g., homophily); the second set of hypotheses center around balance theory \citep{cartwright56,heider58} and its predictions for interaction between two opposing groups; and the last set of hypotheses are constructed via the natural cyclic rhythms of modern society \citep{shumway06}. 

The sampling frame employed by \cite{butts09b} guarantees two distinct groups, specifically DNC designated blogs and RNC designated blogs. In this context these groups represent two contentious factions competing for very real and tangible stakes in the US political arena \cite[see][etc.]{drezner08}. We may view these two groups as halves of an adversarial relationship \citep{hargittai08}, and in doing so may further characterize their interaction through the lens of balance-theoretic notions \citep{cartwright56,heider58}. This allows us to test different influences of dynamic notions of balance: is a DNC blog more likely to cite another DNC blog? is a DNC blog citing an RNC blog less likely to cite an RNC blog at the next time step? and so forth. In this paper we will assume that blogs designated by the DNC represent a cohesive group, blogs designated by the RNC represent a cohesive group, and that the interaction between the two groups represents a negative relation.

\subsubsection{Mixing}

In the social network literature a priori group partitioning in the model and group interaction is often known as \emph{mixing} or \emph{nonrandom mixing}. \cite{hargittai08} hypothesizes that bloggers cluster ideologically and thus will only link to other blogs with the same ideology. We may re-express this hypotheses as a type of assortative mixing process where blogs have an almost exclusive propensity to cite within-group and not across-group. 

A counter hypothesis arises from our a prior grouping and the assumption that these two parties represent competing organizations. A reasonable assumption, given our population, is that between group citations represent a negative relation such that if a blog from the DNC cites a blog from the RNC it is an action performed to refute a claim made by the RNC and visa versa. Under this frame a natural hypothesis arises where we might expect that the propensity to cite across-group will be high and hence represents blog feuding.  The practical application of this hypothesis is that we may then infer that blog citations are primarily negative, and are used to damage their opponent. 

Thus, we propose two contending hypotheses for mixing:

\begin{description}
\item[Mixing Hypothesis 1] The influence of mixing on the utility of a given blog will highly favor in-group mixing and down play out-group mixing.
\item[Mixing Hypothesis 2] The influence of mixing on the utility of a given blog will highly favor out-group mixing and decrease in-group mixing.
\end{description}

\subsubsection{Balance-Theoretic Influences}

\cite{cartwright56} introduced the concept of generalized balance based on \citeauthor{heider58}'s (\citeyear{heider58}) theory of balance, which suggests a number of possible mechanisms for how an actor chooses whether to link between two competing groups in a dynamic context. Heider's theory stems from Gestalt psychology and posits that individuals attempt to bring their system into \emph{balance}. Cognitively, this has the implication that an actor prioritizes links with other individuals of the same attitude, et cetera. The theory of balance suggests a number of different possible mechanisms which might influence an actor's utility function and thus the weights on their payoff elements ($s$). 

We propose the four competing hypotheses (BT Hypothesis 1-4), broken up into two competing sets: BT Hypothesis 1 versus BT Hypothesis 2 and BT Hypothesis 3 versus BT Hypothesis 4.  The first group tests the influence of extended in-group and out-group effects and the second group tests reciprocity effects. \\

\begin{description}
\item[BT Hypothesis 1: Ally of an ally (in-group two paths)] We hypothesize that the chance of a link between two blogs is increased if the edge is embedded in a two-path contained in the same group, (e.g., in this case, the RNC or DNC).  In other words, it is more likely that a relationship will be formed between Ego and an ``ally" if that ally is also connected to a ``ally  (Figure \ref{in-group}).\\

\begin{figure}[h]
\begin{center}
Ego $\rightarrow$ Ally A $\rightarrow$ Ally B $\Rightarrow$ Ego $\rightarrow$ Ally B
\caption{Graphic of in-group two path (ally of an ally).}\label{in-group}
\end{center}Ê
\end{figure}

\item[BT Hypothesis 2: Ally of an opponent (cross-group two paths)] We hypothesize that the chance of a link between two blogs is decreased if the edge is embedded in a two path where the first actor is an ally and the second actor is a hostile (e.g., RNC to DNC, Figure~\ref{crossgroup}).\\

\begin{figure}[h]
\begin{center}
 Ego $\rightarrow$ Ally A $\rightarrow$ Hostile B $\Rightarrow$ Ego $\rightarrow$ Hostile B
\caption{Graphic of cross-group two path (ally of an opponent).}\label{crossgroup}
\end{center}Ê
\end{figure}

\item[BT Hypothesis 3: Reciprocity (ally)]  We hypothesize that reciprocity will be accompanied with positive gains for in-group edge creation and low or negative gains for across-group citation. In this case citations are primarily a positive relation such that utility gain comes from increasing the prominence of one's allies, and not through competition with one's opponents (Figure~\ref{repfriend}). 

\begin{figure}[h]
\begin{center}
Ally $\rightarrow$ Ego $\Rightarrow$ Ego $\rightarrow$ Ally
\caption{Graphic of reciprocity between allies.}\label{repfriend}
\end{center}Ê
\end{figure}

\item[BT Hypothesis 4: Reciprocity (hostile)]  Conversely we hypothesize that reciprocity will be accompanied with positive gains for out-group edge creation and low or negative gains for in-group citation. In this case citations are primarily a negative relation such that utility gain comes from refuting enemies' accusations (Figure~\ref{rephost}).   \\

\begin{figure}[h]
\begin{center}
Hostile $\rightarrow$ Ego $\Rightarrow$ Ego $\rightarrow$ Hostile 
\caption{Graphic of reciprocity between opponents.}\label{rephost}
\end{center}Ê
\end{figure}

\end{description}

\subsubsection{Context and Seasonality}

Networks tend to have certain basic structural characteristics which should be accounted for in any network analysis (e.g., sender and receiver effects; \cite{wasserman.faust:1994a}). 

In a time-series context it is known that there are certain seasonal and period effects that occur in any temporally collected data \citep{shumway06}. Common seasonal effects in behavior data include daily and hourly effects (e.g., Monday, Tuesday, etc. and midnight versus midday). Below are series of temporally motivated hypotheses, which take into account the interaction between structural properties of the network under question and the daily fluctuations of human interaction, such as a differential propensity to update one's links (an effortful procedure) over the course of the day.

\begin{description}

\item[Seasonality Hypothesis 1]\cite{butts09b} found that the volatility of the blog networks changes with time of day, day of week, and period in the electoral cycle.  Translating their notion of ``volatility'' into the present modeling framework, we posit that the degree of inertia in network structure (i.e., the lag effect) varies systematically with time.  

\item[Seasonality Hypothesis 2] We suspect that overall propensity to send links will vary over time. We argue that ego's linking to others involves a search process, and is consumptive of attentional/energetic resources. Resource availability varies, and with it perhaps the total number of links maintained by each blog.

\item[Seasonality Hypothesis 3] We propose that behavioral factors might change with time and context.  Specifically, we hypothesize a mechanism of \emph{selective salience}, in which the propensity to create ties within or across groups increases during important events in the election cycle (see Table~\ref{epochs}).

\end{description}

\section{Methodology}
\label{sec:method}

This work employs the Dynamic Lagged-Logistic Network Regression methodology recommended by \cite{almquist10} for large dynamic data sets, which builds on the Exponential Random Graph \citep{holland81a, holland81b, butts08, snijders06,strauss} and Network Regression literatures \citep{krackhardt87a,krackhardt87b,krackhardt88}.  This model family is particularly appealing in this context because it is very natural to model citation dynamics as a binary choice process, and this framework allows us to explore the mechanisms that predict whether one blogger chooses to cite another blogger. 

All computation for this article was written and executed in the {\tt R} environment \citep{R}. We employ a modified form of the code used in \cite{almquist10}.

We begin by discussing the necessary statistical details needed to connect our inferential framework with our theoretical framework. We then follow this discussion with the operationalized version of the mechanisms discussed in Section \ref{hyp}.

\subsection{Inferential Frame-Work: Dynamic Lagged-Logistic Network Regression}

A standard inferential framework for network analysis is that of the Exponential Family Random Graph Models (ERGMs) \citep[][etc.]{butts08,holland81a}.  Sometimes misconstrued as referring to a narrow class of models, the ERGM framework is better understood as a general approach to the representation of statistical models for network data.  In the case of networks on a fixed set of individuals, we may write the likelihood of observing a given network in exponential family form as

\begin{align}
P_\theta(A=a|X) = \frac{\exp\{\theta^t t(a,X)\}}{c(\theta,X)} \mathbb{I}_\mathcal{A}(a)
\end{align}

\noindent
where $A$ is a random network (represented by its adjacency matrix) on $n$ nodes, drawn from some set $\mathcal{A}$ of potentially observable networks.  (In the present context, $\mathcal{A}$ is the set of all directed networks on the set of RNC and DNC blogs.)  $X$ is then a set of covariates, $\theta$ is a vector of real-valued parameters, $t$ is a vector of graph statistics on $a$ and $X$ (e.g. structural properties and covariate effects), and $\mathbb{I}_\mathcal{A}(a)$ is an indicator function that returns 1 if $a$ is in the set of potentially observable graphs, and 0 otherwise. $c(\theta,X)=\sum_{a'\in \mathcal{A}} \exp\{\theta^t t(a',X)\}$ is simply the sum of the numerator over all observable networks; a normalizing factor, $c(\theta,X)$ ensures that the total probability of all potentially observable graphs sums to 1. 

\cite{almquist10} demonstrate that if one assumes the network only depends upon the past history and/or on exogenous factors (i.e., the covariate set $X$)  one can show that the above model form simplifies to \emph{lagged-logistic network regression}. Drawing on ERG theory, they show that this allows us to write the conditional log-odds of an edge in logistic form, i.e.

 \begin{align}
\textrm{logit}\Pr(A_{ij,t}=1) = \theta^t\left[t_{ij}(A  | A_{ij,t}=1,X)-t_{ij}(A  |  A_{ij,t}=0,X)\right]
\end{align}

In Section \ref{sec:logitchoice} we introduced the notation and theory of a dynamic logistic-choice model. Under the assumptions of a logistic-choice framework (Equations~\ref{eq:logit1}, \ref{eq:logit2}) and the independence assumptions of Section \ref{sec:edgeup} it is clear that in our case the logistic-choice model is indeed lagged-logistic network regression where the inferred parameters represent the weights each blogger places on the elements of his or her payoff function (i.e., utilities).  Thus, we can implement our decision theoretic model for blog dynamics by expressing our hypothesized mechanisms (payoff function components) in terms of model statistics ($t$), and fitting the resulting lagged-logistic regression model to estimate the unknown components of the utility function.

\subsection{The Dynamic Decision Model}

To operationalize our boundedly rational choice model of blog network dynamics, we must express our hypothesized mechanisms in terms of statistics that measure the inputs to each actor's payoff function.  Here, we consider each group of mechanisms, and discuss how they are implemented within the model

\subsubsection{Mixing Terms}

Methods for modeling nonrandom mixing have been known for some time; for our purposes we employ a method similar to \cite{morris91}, and use a type of block model to represent the two groups \cite[see][for a full review of the literature on block modeling]{wasserman.faust:1994a}. In doing so we impose four parameters on the model: one parameter for each group's internal interactions, and one parameter for each group's tendency to send ties to the other. (Note that since this is a directed network RNC$\rightarrow$DNC is different than DNC$\rightarrow$RNC.)  We assume these two groups represent competing organizations and thus expect to see a higher propensity for within group citation than between group citation. 

The mixing terms are modeled as a block matrix, such that the model contains an variable for the number of edges in each of four categories: DNC$\rightarrow$DNC edges, RNC$\rightarrow$RNC edges, DNC$\rightarrow$RNC edges, and RNC$\rightarrow$DNC edges. This set of terms is jointly identifiable, taking the place of the standard edge (count of the number of edges in the model) or density (number of edges divided by the total number of possible edges) terms frequently encountered in ERG models \citep{goodreau.et.al:jss:2008}.

Expressed in the language of mixing terms, our competing interaction hypotheses can be restated as follows:
\begin{description}
\item[Mixing Hypothesis 1] The weights of the in-group effects will be large and positive and cross-group effects will be small or negative; i.e. the model will favor in-group, but not cross-group mixing. 
\item[Mixing Hypothesis 2] The weights of the in-group effects will be smaller than the cross-group effects; i.e. the model will favor cross-group mixing. 
\end{description}

\subsubsection{Heiderian Terms}

Balance Theory is naturally testable in a dynamic network context and may be constructed from classic network decompositions such as two-stars and reciprocal links \cite[see,][]{wasserman.faust:1994a}.  Our balance-theoretic hypotheses can then be restated in operational terms as follows:

\begin{description}
\item[BT Hypothesis 1: Ally of an ally (in-group two paths)] A count of the number of in-group two paths a given edge is involved. We expect the weight on this term to be positive and significant.  
\item[BT Hypothesis 2: Ally of an opponent (cross-Group two paths)] A count of the number of cross-group two paths a given edge is involved. We expect the weight on this term to be negative and significant.  
\item[BT Hypothesis 3 and : Reciprocity (friendly/hostile)] An indicator if a relation is reciprocal and between in-group members and an indicator if a relation is reciprocal and between group members. We expect these terms to have opposite signs (positive and negative for BT 3 and negative and positive for BT4). 
\end{description}

\subsection{Network Effects and Seasonality}

It is often hypothesized that certain basic characteristics of the network understudy are likely to influence payoff elements for each actor in the network. Two often studied structural effects are those of indegree and outdegree \citep[this is sometimes known as preferential attachment, see, ][]{merton68}. We also include a clique comembership term (clustering)\endnote{For algorithmic details see \cite{eppstein10}.}\citep[see, ][]{wasserman.faust:1994a}. We posited above that we might expect the nature of this effect to change dynamically with the the time of day. This may be interpreted as an interaction effect between the seasonality dummy and the indegree term. Below we will discuss the details of the seasonality terms employed in this paper. 

\subsubsection{History and Seasonality}

The effects of past interaction history and seasonality are hypothesized to act as follows:
\begin{description}

\item[Seasonality Hypothesis 1] There will be substantive and large inertial effect, i.e. the lag term will be large and significant.

\item[Seasonality Hypothesis 2] We test the hypothesis that the overall propensity to send links will vary over time via an interaction between hourly fixed effects and  receiver effect and hourly fixed effects and the inertia term. We expect these to be important terms in the model, to be large and significant.

\item[Seasonality Hypothesis 3] 

\begin{description}
\item[Selective salience]  We test the ``selective salience" hypothesis with nine period effects, we expect there to be increase in activity during \emph{PreCon, DNCCon, RNCCon, Deb, Elec} and decrease in activity during \emph{InterCon, PreDeb, PreElec,} and \emph{PostElec} (See Table \ref{epochs} for details).

\end{description}
\end{description}

To model hourly seasonality we employ three dummy variables for each hour ($06$, $12$, and $18$ in the day with hour $0$ as the reference group ($\phi_{06}$, $\phi_{12}$, $\phi_{18}$). 

To model weekly seasonality we employ some ideas from Harmonic Regression \citep{shumway06}.\endnote{Equation \ref{eq1} is used to model daily seasonality (days within the week). We assume a classic ``signal in noise" with a hidden periodic signal. If we assume there is a single sinusoid,  we can model the weekly cycles as follows
\begin{align*}
R\cos(2\pi \omega_d t+\Phi)
\end{align*}
Using the classic trigonometric formula $\cos(a + b) = \cos(a) \cos(b)-\sin(a)\sin(b)$ we can derive the terms we place in the model (Eq. \ref{eq1}). 
\begin{align}
R\cos(\Phi)\cos(2\pi\omega_d t)+ - R\sin(\Phi)\sin(2\pi\omega_d t) \nonumber \\
\theta_1 \cos(2\pi\omega_d t)+\theta_2 \sin(2\pi\omega_d t). \label{eq1} 
\end{align}
} This amounts to assuming the influence of a given week in a month is cyclic (see Section~\ref{sec:find} for more details).

\section{Analysis}

We employ standard model selection techniques (selection by BIC) and adequacy choices to determine the mechanisms that appear to be active in shaping actors' choice process; after identifying the relevant mechanisms, we examine the parameters of the best fitting model to interpret the implied utility function. A natural way to begin our analysis is to first construct a simple, baseline model which contains only the mixing terms and seasonal effects (Model 1). We then add a single lag term (Model 2), followed by network control effects (Model 3). Next, we add in the Heiderian mechanisms (Model 4); lastly we add in period effects (Model 5), and hourly interaction terms with sender and inertia (Model 5).  The parameter estimates and BIC scores for each of these models may be found in Table~\ref{models}.

\begin{center}
[Table~\ref{models} About Here ]
\end{center}

\subsection{Model Adequacy Check}

We start by employing the BIC model selection criterion \citep{schwarz78}, where the model with the lowest BIC is chosen as the preferred model. The relative performance of our candidate models provides us with the first evidence regarding our hypotheses: if a given effect does not appear in the preferred model, then the data suggests that the associated mechanism is not influential in the actors' choice processes.  Hypothesized effects not found in the best fitting model are thus rejected.

Although likelihood-based model selection criteria such as the BIC are effective in evaluating the relative performance of competing models within a specified set, they do not evaluate the \emph{adequacy} of the selected model in substantive terms.  To verify that our selected model is adequately able to reproduce the basic features of our data set, we follow model selection with simulation-based adequacy assessment \citep{almquist10,hunter08}. The method employed is as follows.  First, we choose a set of Graph Level Indices (GLI) \citep{anderson99} capturing various structural features of the blog networks.  Given this set of network measures, we use the best-fitting model at each time step to simulate (forecast) the blog network to be observed at the next time step.  The distribution of the network properties under the predicted networks are then compared to the values actually observed.  While we do not expect to perfectly forecast network evolution, we expect our model to produce predictions which are generally compatible with the evolving data.

\begin{sidewaysfigure}
   \centering
   \includegraphics[width=1\linewidth]{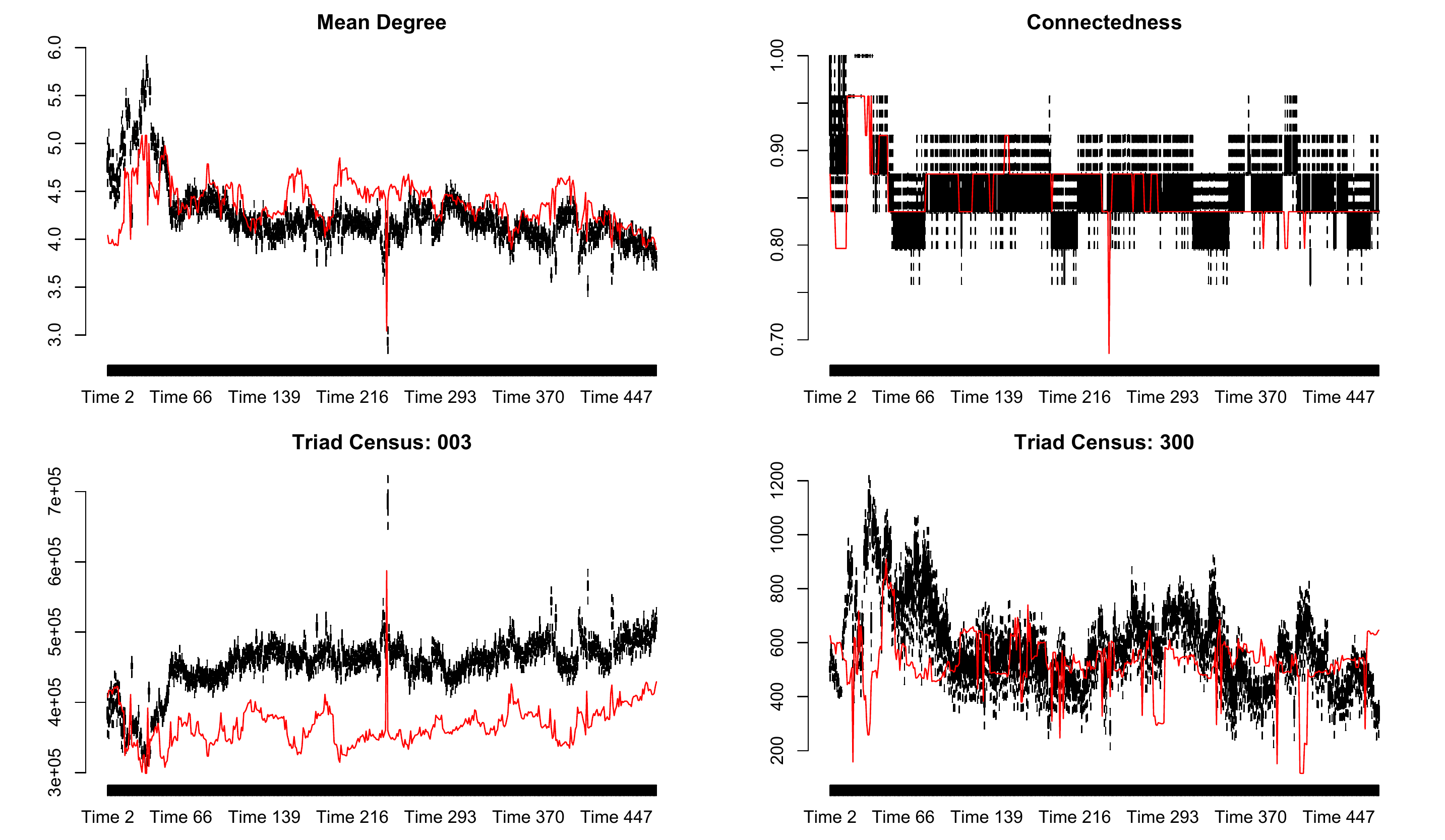} 
   \caption{Graph Level index  comparison for one-step one-lag network logistic regression for Model 5 with 100 simulations at each time point.}
   \label{fig:glionestep}
\end{sidewaysfigure}

For present purposes, we evaluate model performance with respect to the following graph-level indices: density \citep[known to be a very important graph statistic, see][]{wasserman.faust:1994a}; Krackhardt's connectedness index\citep[a measure of reachability and to what extent information might flow through the network, see][]{krackhardt94}; mean indegree (receiver) and mean outdegree (sender) \citep[a common graph statistic, involved in theories of preferential attachment (receiver) and expansiveness (sender), see][]{wasserman.faust:1994a}; the ``null'' or ``empty'' triad \citep[designated as type 003 in Holland and Leinhardt's typology a measure of isolates or actors not engaged in a given period, see][]{wasserman.faust:1994a}; and lastly the number of triangles or cliques \citep[the complete triad, 300 in Holland and Leinhardt's typology, an indicator of clustering or local interaction, see][]{faust10}. The realized and forecast values for these statistics are shown in Figure~\ref{fig:glionestep}.

\begin{center}
[Figure \ref{fig:glionestep} About Here ]
\end{center}

We can see from Figure~\ref{fig:glionestep} that Model 5 is very effective in capturing the basic trend for density, mean indegree/outdegree, triangles, and connectedness (Figure~\ref{fig:glionestep} and Table~\ref{models}), with observed values generally close to or within the forecast interval.  The model forecast for null triads also tracks the trend, although the predicted value is somewhat higher than the observed.  Overall, the adequacy checks suggest that the model captures the main features of the blog dynamics, and we therefore move forward with our analysis.

\subsection{Findings}
\label{sec:find}

Parameter estimates for Model 5 are shown in Table~\ref{models}.  It is important when interpreting the parameters of Model 5 to recall that many effects are necessarily simultaneous, and should be viewed as a group. Take, for example, the mixing terms: while two of the mixing terms are negative, they cannot be interpreted without taking into account the lag term, which is larger than any one of the mixing terms and positive.  Thus, while citation is costly in general, there is still a net tendency to cite those one has cited in the previous time step.  We see that the base propensity for within group blog-to-blog citation is higher for the RNC than DNC, and that the two groups are about equally likely to send cross-group ties (on an edgewise basis). 

In terms of our hypotheses, we confirm Mixing Hypotheses 1 that within group linking is more likely than between group linking and refute Mixing Hypothesis 2 as the two groups have similar levels of between group linking. Payoffs to in-group citations are higher than cross-group citations; noting that this is less true for the RNC on a per-tie basis.

We confirm our balance-theoretic hypothesis of ally-of-ally connections; two-path embeddedness act as citation incentives (Table~\ref{models}: Group-2-Path, Cross-Group-2-Path). One important implication of this result is that citations are more likely to ``flow'' through allies or competitors than across.  In contrast with this result, we find that reciprocity does \emph{not} follow a classic balance-theoretic pattern.  Ceteris paribus, incoming ties from one's own allies tend to \emph{reduce} the propensity of a returning citation, while those of one's opponents \emph{increase} this propensity (Table~\ref{models}: Group-Reciprocity, and Between-Group-Reciprocity). Although inconsistent with a positive, exchange-theoretic notion of citation, this is compatible with the notion that (1) citation within groups is redundant (or event deferential), and hence poorly reciprocated, while (2) citation between groups tends to take the form of critique or conflict, in which ``returning fire'' is highly incentivized.  This finding underscores the importance of considering group or institutional context when theorizing the nature and role of reciprocating mechanisms.

\begin{figure}[htbp]
   \centering
   \includegraphics[width=1\linewidth]{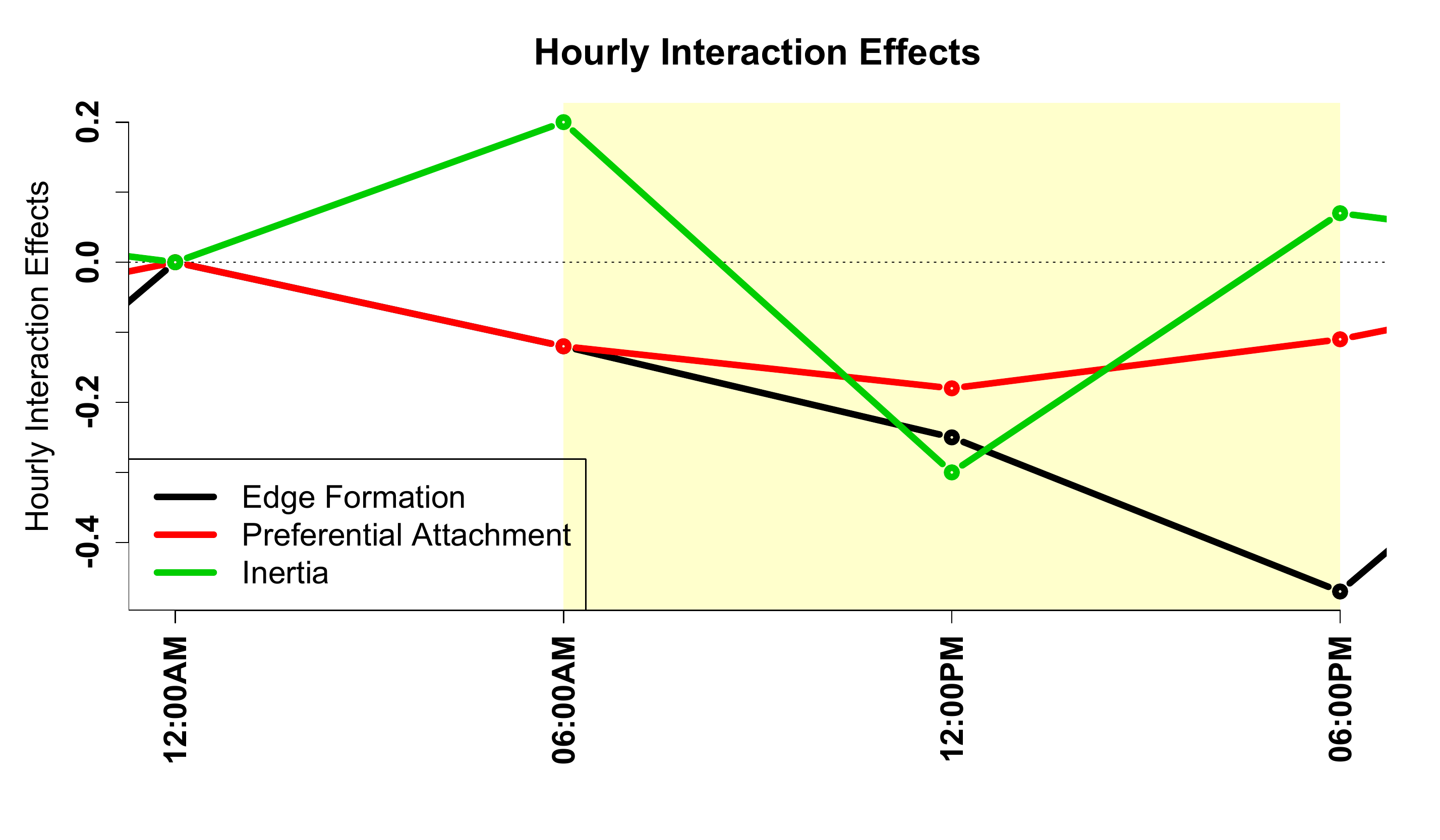}
   \caption{Visualization of the interaction edge formation payoff effects over a 24 hour period.}
   \label{fig:hourlyInter}
\end{figure}

\begin{figure}[htbp]
   \centering
   \includegraphics[width=1\linewidth]{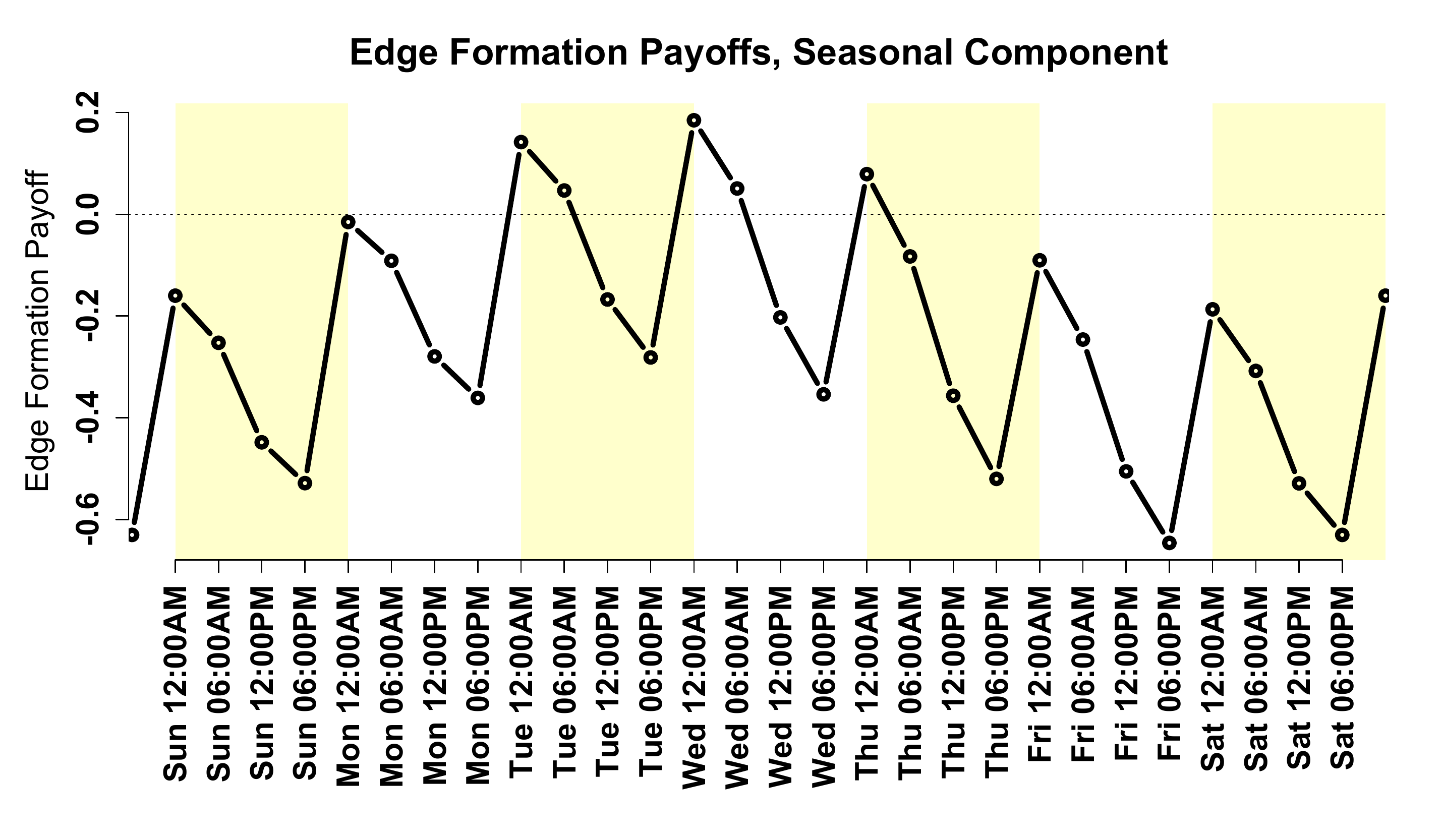}
   \caption{Visualization of the baseline edge formation payoffs by over one week.}
   \label{fig:formation}
\end{figure}

We confirm Seasonality Hypothesis 1, finding that inertia is a strong and persistent effect in this network. To interpret the baseline payoff effects of the seasonal components and their interaction effects we have plotted the hourly and weekly baseline (Figure~\ref{fig:formation} and ~\ref{fig:hourlyInter}). Figure~\ref{fig:formation} demonstrates that the baseline propensity to form an edge evolves in a systematic and periodic fashion as a superposition of daily and weekly cycles. The propensity to form ties to others is highest at the start of the day, and declines as the day goes on; the effect also builds and recedes during the week, and is at its lowest during the weekend. Figure~\ref{fig:hourlyInter} allows us to see how other mechanisms interact with daily seasonality: inertia is at its highest in the early/mid-morning, and in the early evening. Sensitivity to others' popularity is at its highest at night (when tie formation tendencies are also their strongest); this suggests that bloggers' activities do not reflect a uniform pattern of behavior, but instead show a pattern of \emph{attention shift} tied to daily and weekly activity patterns.  The specific pattern observed is consistent with an ``information processing cycle,'' in which actors begin the day by reactively posting ties to other blogs widely attended to by peers, then shift into a period of media consumption (and low posting activity), a period of link purging and posting ties to new primary sources (rather than other blogs), and finally a period of quiescence before the cycle begins anew.  This regular shift in the nature of activity leads to a network that does not evolve in a uniform, even manner, but that instead ``pulses'' as new information is drawn in, old ties are removed, and new ties are formed in response to the day's events.  Such a pattern is a marked departure from the temporally uniform behavior assumed by most current models of group dynamics.

\begin{center}
[ Place Figure~\ref{fig:formation} About Here ]
\end{center}

\begin{center}
[ Place Figure~\ref{fig:hourlyInter} About Here ]
\end{center}

Lastly, we obtain mixed results for Selective salience hypothesis (calling that we have to interpret the associated terms in relation to the lag term). We see the propensity to form ties is greatest during the PreCon and DNCCon periods, being generally stable during much of the rest of the electoral cycle (the exception being election day, when there is a noticeable decrease in edge creation).  We thus see little to no evidence of ``important'' events fostering edge indication, and some suggestion that very important events (e.g. the election) may reduce citations within the blog network.  Interpreted in terms of the ``information processing cycle'' described above, this may reflect increasing emphasis during such periods on attention to external sources, and a concomitant reduction in intra-network activity.

\section{Discussion and Conclusion}

In this work we have modeled the evolution of an online political interaction network as a logistic choice process, treating blog authors as boundedly rational actors engaged in choices regarding their outgoing citations \cite[similar in concept to][]{snijders01}. To implement this model, we have employed a type of lagged logistic network regression \cite[see][]{almquist10} that arises as a natural consequence of our assumed decision making process.  Drawing on existing ideas from the group behavior literature, we have identified a number of candidate mechanisms of potential relevance to the choice process.  By fitting our decision model to the observed network data, we have been able to determine which of these mechanisms do, in fact, seem to influence individual choices, and to determine the nature of the influences in question. The result provides us with a composite picture of social behavior that, while familiar in many respects, highlights the need for a deeper conceptualization of the role of context effects -- particularly temporal context -- on group dynamics.

Among the anticipated findings, we see that our two groups of respective allies (the RNC and DNC credentialled blogs) have a greater propensity for in-group citation than between-group citation (consistent with \citet{hargittai08}). We also find a tendency towards triadic closure within groups (the ``ally of an ally'' mechanism), as expected on balance theoretic grounds. On the other hand, triadic closure is also observed across groups, an effect that is not anticipated by balance theory.  Likewise, reciprocity dynamics behave quite differently than balance would suggest: although I reciprocate ties from opponents, I tend \emph{not} to reciprocate ties from allies, as would be the case if ties had uniformly positive implications. A potential resolution of these otherwise puzzling effects can be found in the differing nature of within-group and cross-group interaction. Cross-group interaction in our setting is generally rivalrous, and reciprocity generally an act of self-defense (with the ``ally-of-an-opponent" akin to an act of ``piling on'' to an opponent attacked by an ally). On the other hand, within-group citation appears to have a more functional role, and may reflect an underlying hierarchy of information dissemination (accounting for both lower reciprocity and a tendency towards transitive closure).  Fundamentally, our findings are consistent with the observation that \emph{the social meaning of a tie is dependent on group and/or institutional context, and that the dynamics of network evolution are sensitive to these distinctions.}  Developing a richer set of theoretical propositions to predict precisely when, and how, these distinctions will be made would seem to be an important topic for future work.

An even more basic lesson of our findings, however, is that \emph{individual behavior and group dynamics are not uniform, but governed by regular cycles that affect both the nature and extent of activity.}  While we are used to thinking of non-human animals as being governed by diurnal and seasonal cycles, sociologists have been slow to recognize the role of cyclic influences on choice behavior and the dynamics of interaction.  While theorists as wide-ranging as \citet{sorokin:bk:1957}, \citet{mayhew:sf:1980}, and \citet{elder:bk:1974} have argued for greater attention to the role of temporal and environmental context as determinants of social outcomes, these calls are at best poorly reflected in current sociological research.  Our findings suggest the seriousness of this gap: without considering temporal context, one is at a loss to account not only for individual choices, but for the evolution of the social system as a whole.  To the extent that phenomena such as the ``information processing cycle'' posited here hold in other settings, omission of seasonal effects likewise hinders our ability to understand the functional characteristics of social systems, and thus to anticipate the effects of planned or unplanned interventions.  Even online, the physical realities of daily life and the institutional settings in which persons and groups are embedded provide a powerful and dynamic influence on social evolution

\theendnotes

\bibliographystyle{ajs}
\bibliography{blog}

\clearpage

 \begin{sidewaystable}
\begin{tabular}{llrrr}
\hline
\hline
Epoch & Description & Start Time & End Time &Time Points \\
\hline
PreCon & Start of window to DNC Convention & 7/22, 00:00 & 7/25,18:00 & 16 \\
DNCCon & DNC Convention & 7/26, 00:00 & 7/29, 18:00 & 16\\
InterCon & End of DNC Convention to start of RNC Convention & 7/30, 00:00 & 8/29,18:00 & 124\\
RNCCon & RNC Convention & 8/30, 00:00&  9/2, 18:00 & 16 \\
PreDeb & End of RNC Convention to first presidential debate & 9/3, 00:00 & 9/20, 12:00 & 71\\
Deb & First presidential debate to last presidential debate & 9/20, 18:00 & 10/14, 18:00 & 93 \\
PreElec & Post last presidential debate to Election Day & 10/14, 00:00 & 11/1, 18:00 & 76 \\
Elec & Election Day & 11/2, 00:00 & 11/2, 18:00 & 4\\
PostElec & Post election to end of window & 11/3, 00:00 & 11/19, 18:00 & 67 \\
\hline
\hline
\end{tabular}
\caption{Epochs in the 2004 Election Cycle from \cite{butts09b}.}\label{epochs}
\end{sidewaystable}

 \begin{sidewaystable}
\begin{center}
\begin{tabular}{rlllll}
  \hline
 & Model 1 & Model 2 & Model 3 & Model 4 & Model 5 \\ 
  \hline
BIC & 598094.826 & 42209.8184 & 41863.9425 & 41246.5485 & 40310.1217 \\ 
 DNC & -1.7960* & -5.8729* & -5.9459* & -6.0266* & -4.5198* \\ 
 RNC & -0.9401* & -4.0849* & -4.1857* & -4.4320* & -3.2081* \\ 
  DNC$\rightarrow$RNC & -3.9527* & -6.9714* & -6.9960* & -6.9992* & -5.4081* \\ 
   RNC$\rightarrow$DNC & -3.3602* & -5.2977* & -5.4960* & -5.7088* & -4.5039* \\ 
  $A_{t-1}$  &  & 10.8933* & 10.7390* & 10.4363* & 10.3164* \\ 
   Cluster   &   &    &                  &                  & \ 0.3402* \\ 
  Receiver  &  &  & \ 0.0633* & \ 0.0353* & \ 0.0470* \\ 
    Sender  &  &  & -0.0523* & -0.0321* & -0.0231* \\ 
  Group-2-Path &  &  &  & \ 0.2165* &\ 0.2125* \\ 
  Cross-Group-2-Path &  &  &  & \ 0.5393* &  \ 0.3316* \\ 
  Group-Reciprocity  &  &  &  & -0.5401* &-0.5108* \\ 
  Between-Group-Reciprocity&  &  &  & \ 0.3752 &  \ 0.6299* \\ 
  $\theta_{1}$ & \ 0.0016 \ & \ 0.0604* & \ 0.0580* & \ 0.0525* &\ 0.1048* \\ 
  $\theta_{2}$& \ 0.0363* & \ 0.1593* & \ 0.1450* & \ 0.1228* &  -0.1584* \\ 
  $\phi_{06}$& -0.0933* & -0.5516* & -0.5128* & -0.4488* &  -0.1209* \\ 
  $\phi_{12}$ & -0.0922* & -0.5899* & -0.5507* & -0.4841* &-0.3459* \\ 

  $\phi_{18}$ & -0.0936* & -0.6887* & -0.6496* & -0.5791* & -0.4670* \\ 

  Receiver$\times \phi_{06}$&  &  &  &  & -0.1169* \\ 
  Receiver$\times \phi_{12}$&  &  &  &  & -0.1826* \\ 
  Receiver$\times \phi_{18}$ &  &  &  &  & -0.1102 \ \\ 
 $A_{t-1}\times \phi_{06}$  &  &  &  &  & \ 0.2045* \\ 
   $A_{t-1}\times \phi_{12}$&  &  &  &  & -0.2980* \\ 
 $A_{t-1}\times \phi_{18}$ &  &  &  &  & \ 0.0704 \\ 
  DNCCon &  &  &  &  & -1.9237* \\ 
  InterCon &  &  &  &  & -1.9549* \\ 
  RNCCon &  &  &  &  & -1.9511* \\ 
  PreDeb &  &  &  &  & -1.8196* \\ 
  Deb &  &  &  &  & -1.8898* \\ 
  PreElec &  &  &  &  & -1.9294* \\ 
  Elec &  &  &  &  & -2.4212* \\ 
  PostElec &  &  &  &  & -1.9480* \\ 
   \hline
\end{tabular}
\caption{Five models for dynamic logistic choice of Inter and Intra-group Blog Citation networks in the 2004 US Presidential Election ordered by BIC. Significant at ``*" 0.05 p-value level under a z-test unless otherwise specified.}\label{models}
\end{center}
\end{sidewaystable}

\clearpage

\end{document}